\documentclass[Physsubmission, Phys]{SciPost}

\hypersetup{
    colorlinks,
    linkcolor={red!50!black},
    citecolor={blue!50!black},
    urlcolor={blue!80!black}
}

\usepackage[bitstream-charter]{mathdesign}
\DeclareSymbolFont{usualmathcal}{OMS}{cmsy}{m}{n}
\DeclareSymbolFontAlphabet{\mathcal}{usualmathcal}

\usepackage{amsmath,graphicx,todonotes,xspace,cite,hyperref,slashed,subcaption,float,tabularx}

\newcommand{\as}{\ensuremath{\alpha_s}\xspace}
\newcommand{\HEJ}{{\tt HEJ}\xspace}

\newcommand{\HEJtwo}{{\tt HEJ2}\xspace}

\begin{document}

\begin{center}{\Large \textbf{
High-energy logarithmic corrections to the QCD component of same-sign W-pair production\\
}}\end{center}

\begin{center}
Jeppe R.~Andersen\textsuperscript{1},
Bertrand Duclou\'e\textsuperscript{2$\star$},
Conor Elrick\textsuperscript{2},
Andreas Maier\textsuperscript{3},
Graeme Nail\textsuperscript{2} and
Jennifer M.~Smillie\textsuperscript{2}
\end{center}

\begin{center}
{\bf 1} Institute for Particle Physics Phenomenology, University of Durham,
Durham, DH1 3LE, UK
\\
{\bf 2} Higgs Centre for Theoretical Physics, University of Edinburgh, Peter Guthrie Tait Road, Edinburgh, EH9 3FD, UK
\\
{\bf 3} Deutsches Elektronen-Synchrotron DESY, Platanenallee 6, 15738 Zeuthen, Germany
\\
* bertrand.ducloue@ed.ac.uk
\end{center}

\begin{center}
\today
\end{center}


\definecolor{palegray}{gray}{0.95}
\begin{center}
\colorbox{palegray}{
  \begin{tabular}{rr}
  \begin{minipage}{0.1\textwidth}
    \includegraphics[width=22mm]{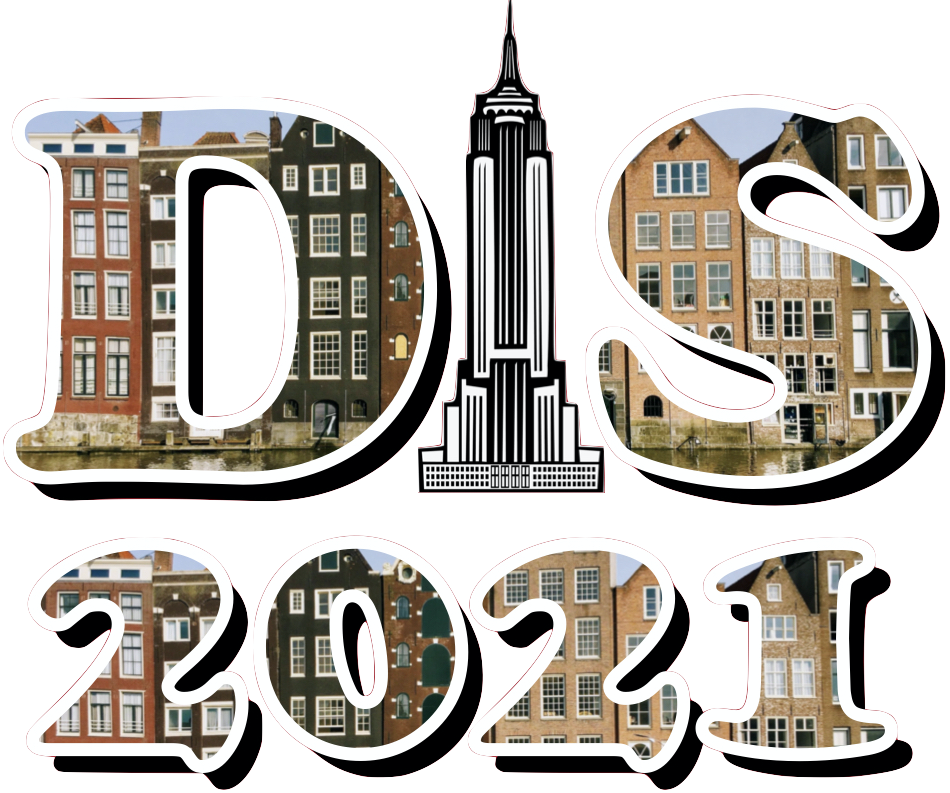}
  \end{minipage}
  &
  \begin{minipage}{0.75\textwidth}
    \begin{center}
    {\it Proceedings for the XXVIII International Workshop\\ on Deep-Inelastic Scattering and
Related Subjects,}\\
    {\it Stony Brook University, New York, USA, 12-16 April 2021} \\
    \doi{10.21468/SciPostPhysProc.?}\\
    \end{center}
  \end{minipage}
\end{tabular}
}
\end{center}

\section*{Abstract}
{\bf
We describe the calculation of the QCD contribution to same-sign $W$-pair
production, $pp\to e^\pm \nu_e \mu^\pm \nu_\mu jj$, resumming all contributions scaling as $\alpha_W^4 \alpha_s^{2+k}\log^k(\hat s/p_\perp^2)$~\cite{Andersen:2021vnf}. These leading logarithmic contributions are enhanced by typical cuts used for Vector Boson Scattering (VBS) studies. We show that while the cross sections are little affected by these corrections, other more exclusive observables relevant for experimental studies are affected more significantly.
}

\section{Introduction}

Vector boson scattering (VBS) represents a key test of the standard model and has been the topic of many experimental and theoretical studies (see e.g.~\cite{Covarelli:2021gyz} for a recent review). In addition to the electroweak (EW) contribution of interest for such studies, other topologies may contribute to the same final state $pp\to VV+2j$. This includes in particular the QCD contribution which starts at $\mathcal{O}(\alpha_W^2 \alpha_s^2)$. To try to increase the relative importance of the EW contribution compared to the QCD one, analyses typically implement so-called VBS cuts which require a large rapidity separation and/or invariant mass between the leading jets in the event~\cite{Aaboud:2019nmv,Sirunyan:2020gyx,Sirunyan:2020gvn}. It turns out that this is precisely the region where logarithmic corrections in $\log(\hat s/p_t^2)$ become important and need to be resummed to all orders since in these kinematics the combination $\alpha_s \log(\hat s/p_t^2)$ is not necessarily small. Therefore, while the VBS cuts are effective at increasing the EW fraction, they also make it more difficult to accurately evaluate the remaining QCD contribution. Here we will discuss how these high energy contributions can be resummed to leading logarithmic (LL) accuracy in the \HEJ framework.

\section{Construction of the Amplitudes}

\begin{figure}
	\centering
	\includegraphics[width=0.65\textwidth]{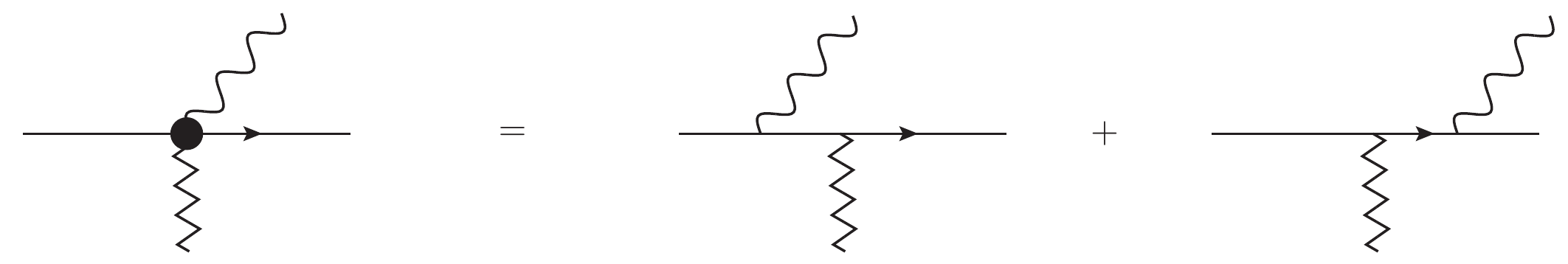}
	\caption{Schematic illustration of the current $j^W_\mu(p_i,p_o,p_\ell,p_{\bar\ell})$ defined in Eq.~\eqref{eq:jW} which describes the production of a $W$ boson from a quark line with an off-shell gluon (shown as a zigzag line).}
	\label{fig:jW}
\end{figure}

In this work we focus on same-sign $W$-pair production, $pp\to W^\pm W^\pm +\ge2j$, where one $W$ boson decays in the electron channel and the other one in the muon channel. 
We start with the leading order (LO) process:
\begin{align}
\label{eq:momww}
q(p_a)Q(p_b) \to (W_1^\pm\to) \ell(p_{\ell_1}) \bar\ell(p_{\bar\ell_1})
(W_2^\pm\to) \ell'(p_{\ell_2}) \bar\ell'(p_{\bar\ell_2}) q'(p_1) Q'(p_2),
\end{align}
with $q$ and $Q$ representing different quark or anti-quark flavours. There are 8 Feynman diagrams contributing at this order, corresponding to the possible ways to assign the $W$s to the different quark lines. To describe the emission of a $W$ boson from a quark line, we define the following current:
\begin{align}
\label{eq:jW}
\begin{split}
j^W_\mu(p_i,p_o,p_\ell,p_{\bar\ell}) =&\ \frac{g_W^2}{2}\ \frac{1}{(p_\ell +
	p_{\bar\ell})^2-m_W^2+i \Gamma_W m_W}\ \left[ \bar u^-(p_\ell) \gamma_\alpha
v^-(p_{\bar\ell}) \right] \\
&\times  \left( \frac{ \bar u^-(p_o) \gamma^\alpha
	(\slashed{p}_o+\slashed{p}_\ell + \slashed{p}_{\bar\ell}) \gamma_\mu
	u^-(p_i)}{(p_o + p_\ell + p_{\bar\ell})^2} +\frac{ \bar u^-(p_o) \gamma_\mu
	(\slashed{p}_i -\slashed{p}_\ell - \slashed{p}_{\bar\ell}) \gamma^\alpha
	u^-(p_i)}{(p_i - p_\ell - p_{\bar\ell})^2} \right),
\end{split}
\end{align}
as schematically represented in Fig.~\ref{fig:jW}. This current is in fact the same as the one appearing in single $W$ plus jets production~\cite{Andersen:2012gk,Andersen:2020yax}.
The exact LO amplitude for the process~(\ref{eq:momww}) is implemented in the \HEJ event generator as a contraction of two such currents.
In order to achieve leading-logarithmic accuracy, this must be supplemented with higher order corrections starting at $\mathcal{O}(\as^3)$. Using the methods of the \HEJ framework~\cite{Andersen:2009nu}, this takes the form of Lipatov vertices
describing real gluon emissions between the two extremal partons and exponential factors for $t$-channel propagators, arising from the virtual corrections and the integration of unresolved real emissions. We refer to~\cite{Andersen:2021vnf} for more details. These corrections capture the leading logarithmic behaviour at high energy. Let us stress that while some approximations are made within the matrix element beyond leading order, the phase space is integrated over in an exact way using Monte Carlo techniques, which gives full flexibility to implement the same cuts as experimental analyses.

While the amplitude as described above already has the correct LL behaviour, its accuracy can be further improved by supplementing it with subleading terms providing LO accuracy at each order in $\alpha_s$ up to the point where this is computationally feasible. This fixed order matching is implemented using the methods of \HEJtwo~\cite{Andersen:2018tnm,Andersen:2019yzo} which is available at \url{https://hej.hepforge.org}. In practice we have used Sherpa~\cite{Bothmann:2019yzt} to generate the fixed-order input for this study, including LO contributions up to 6 jets.

In addition, we also rescale the \HEJtwo predictions to match the inclusive cross section obtained at NLO. However, as explained in the following section, the inclusive cross sections are very close in this particular setup and the matching therefore has a negligible impact.

\section{Impact of the LL Corrections}
\label{sec:results}

We now turn to the \HEJtwo predictions adapted to the 13 TeV LHC kinematics. The cuts on the jets and leptons are the same as in a recent CMS analysis of this process~\cite{Sirunyan:2020gyx}. In addition to the inclusive cuts, CMS also considered additional so-called VBS cuts to try to increase the EW/QCD ratio: $m_{j_1j_2}> 500$~GeV, $|\Delta \eta_{j_1j_2}|> 2.5$ and $\max(z_l) < 0.75$, where $z_l$ is the Zeppenfeld variable~\cite{Rainwater:1996ud}:
\begin{align}
\label{eq:zepp}
z_l = \frac{\eta_l -\frac12 (\eta_{j_1}+\eta_{j_2})}{|\eta_{j_1}-\eta_{j_2}|}.
\end{align}

\begin{figure}
	\centering
	\begin{subfigure}{0.49\textwidth}
		\includegraphics[width=\textwidth]{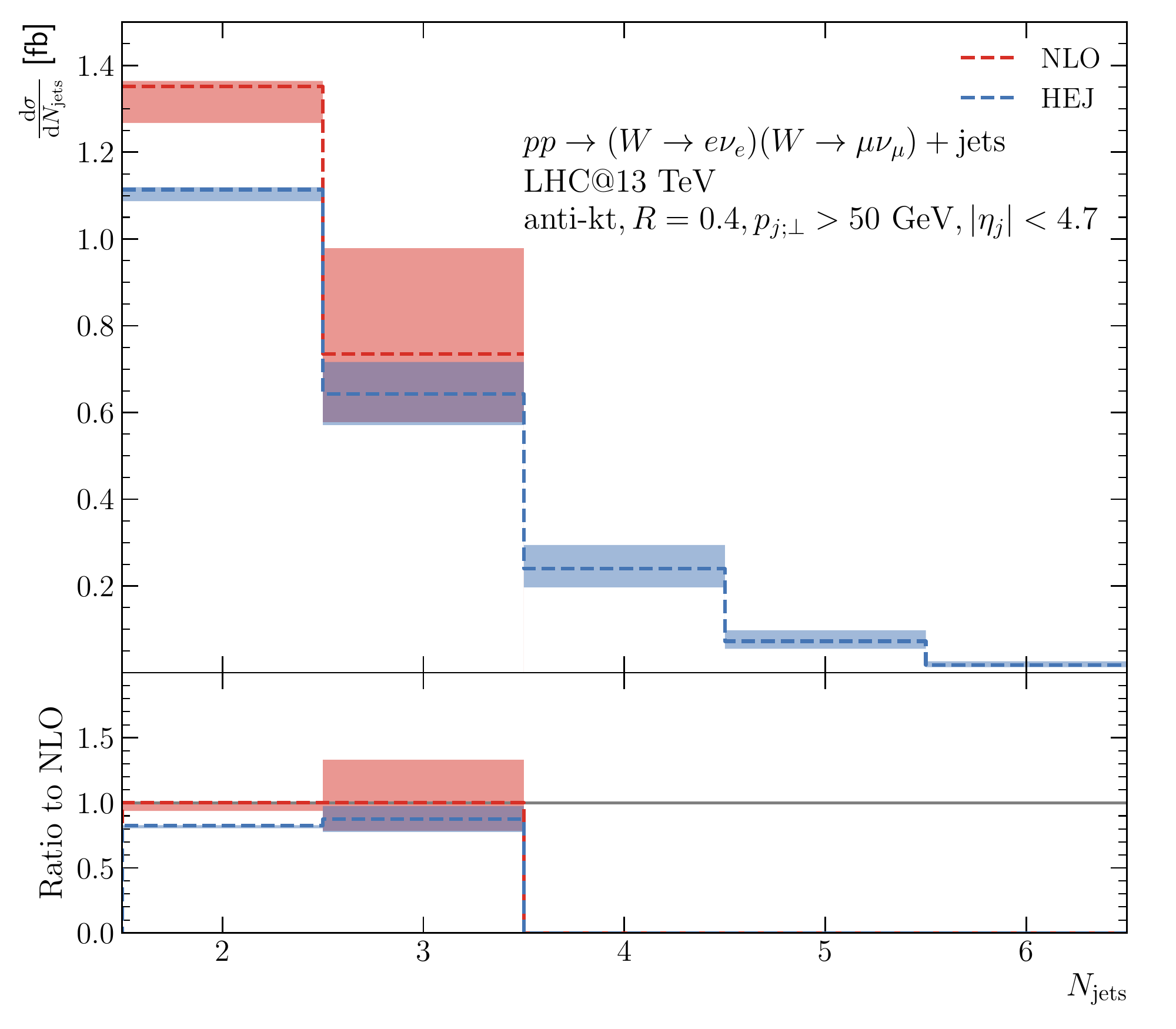}
		\caption{}
		\label{fig:exclusive_jet_rates_novbs}
	\end{subfigure}
	\begin{subfigure}{0.49\textwidth}
		\includegraphics[width=\textwidth]{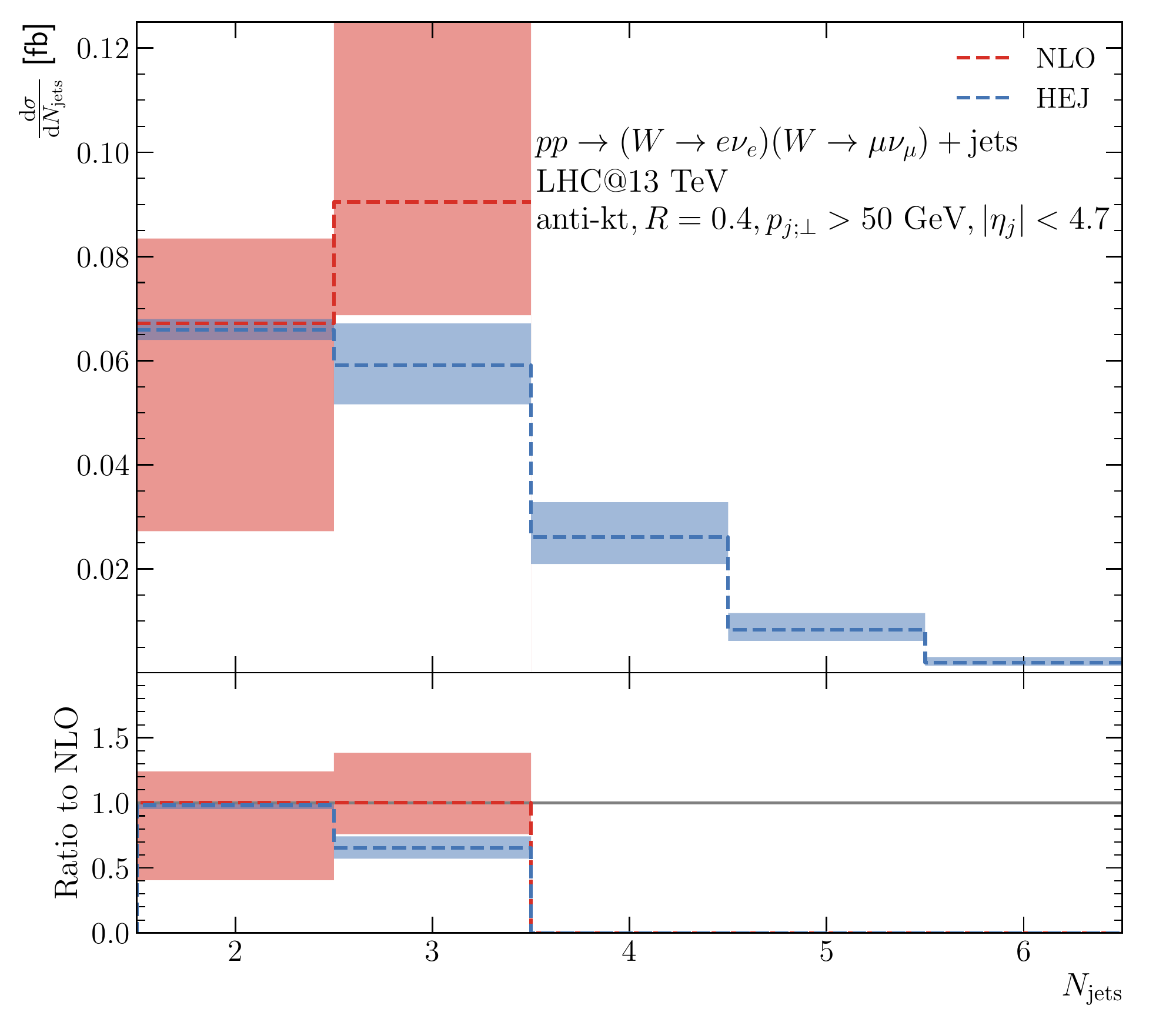}
		\caption{}
		\label{fig:exclusive_jet_rates_vbs}
	\end{subfigure}
	\caption{Exclusive jet rates for $pp\to W^\pm W^\pm +\ge2j$, without (a) and with (b) the additional VBS cuts.}
	\label{fig:exclusive_jet_rates}
\end{figure}

In Table~\ref{tab:crosssections} we show the comparison of the cross sections obtained with \HEJtwo and at NLO using Sherpa~\cite{Bothmann:2019yzt} with the extension of OpenLoops~\cite{Buccioni:2017yxi}. We observe a very close agreement between the two calculations, both before and after the VBS cuts. However, this agreement does not hold for less inclusive observables as can be seen for example in Fig.~\ref{fig:exclusive_jet_rates} where we show the exclusive jet rates obtained both before and after the additional VBS cuts. In the inclusive case one can observe a steady decrease at each multiplicity. On the other hand, after imposing the VBS cuts, the relative importance of the higher multiplicities is significantly enhanced, with the $2$-jet and $3$-jet rates being
very similar in the \HEJtwo predictions, while the $3$-jet rate is even larger than the $2$-jet one in the NLO calculation. These results show the importance of the higher-order corrections resummed by \HEJtwo.

\begin{figure}
	\centering
	\begin{subfigure}{0.49\textwidth}
		\includegraphics[width=\textwidth]{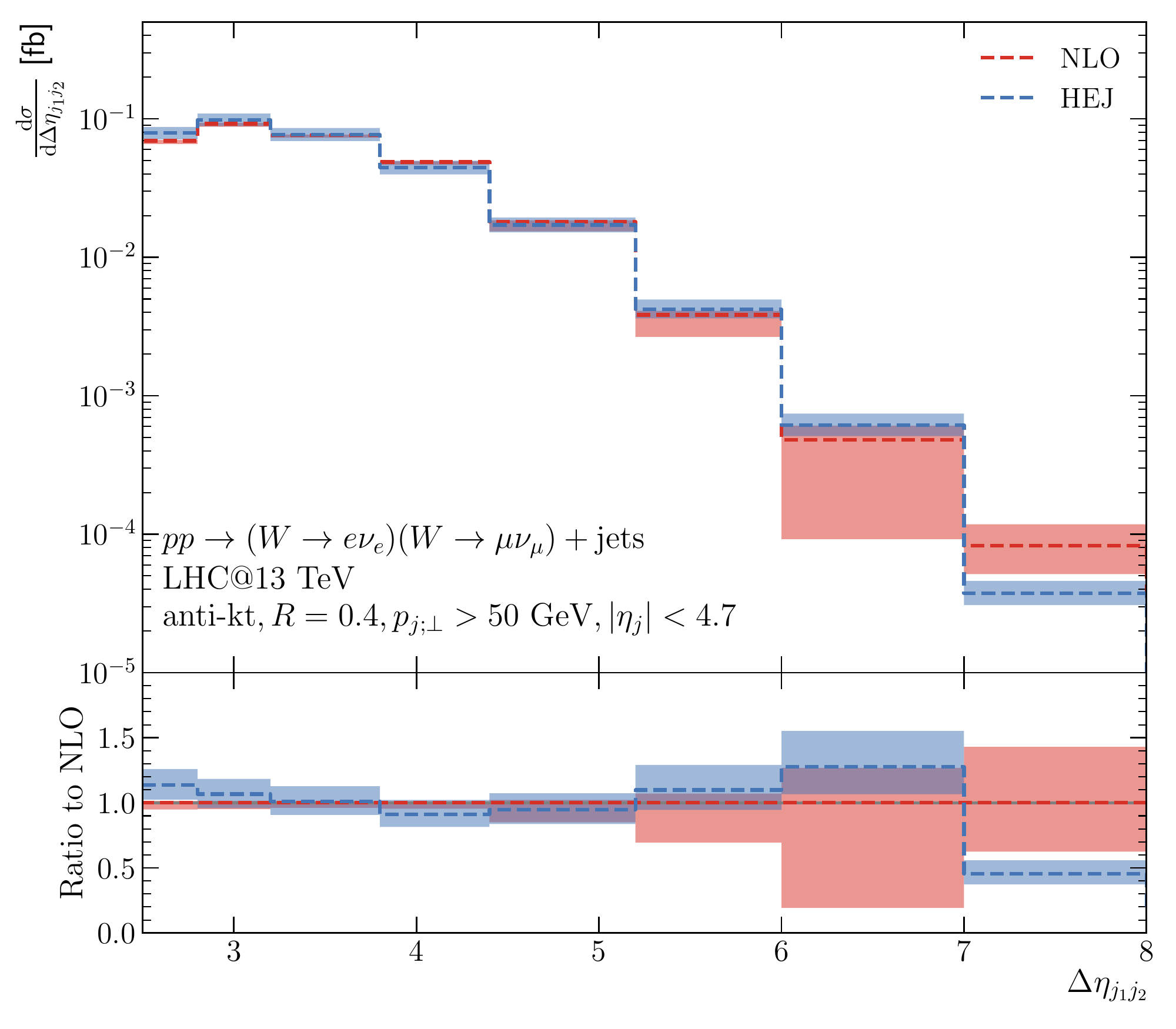}
		\caption{}
		\label{fig:deltay_jj_novbs}
	\end{subfigure}
	\begin{subfigure}{0.49\textwidth}
		\includegraphics[width=\textwidth]{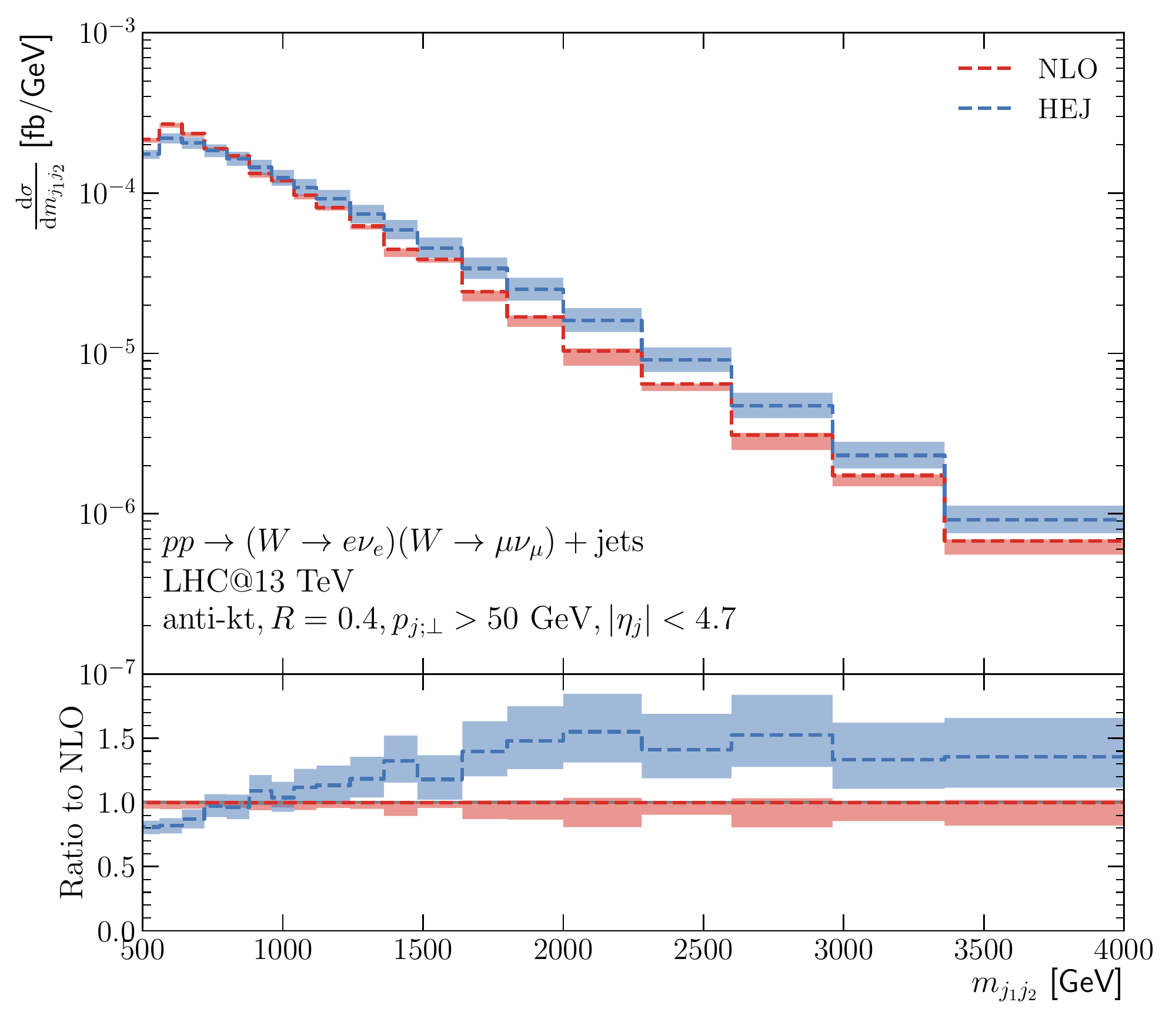}
		\caption{}
		\label{fig:deltay_jj_vbs}
	\end{subfigure}
	\caption{Differential distributions in the pseudorapidity separation (a) and invariant mass (b) of the two leading jets.}
	\label{fig:deltayjj}
\end{figure}

\begin{figure}
	\centering
	\begin{subfigure}{0.49\textwidth}
		\includegraphics[width=\textwidth]{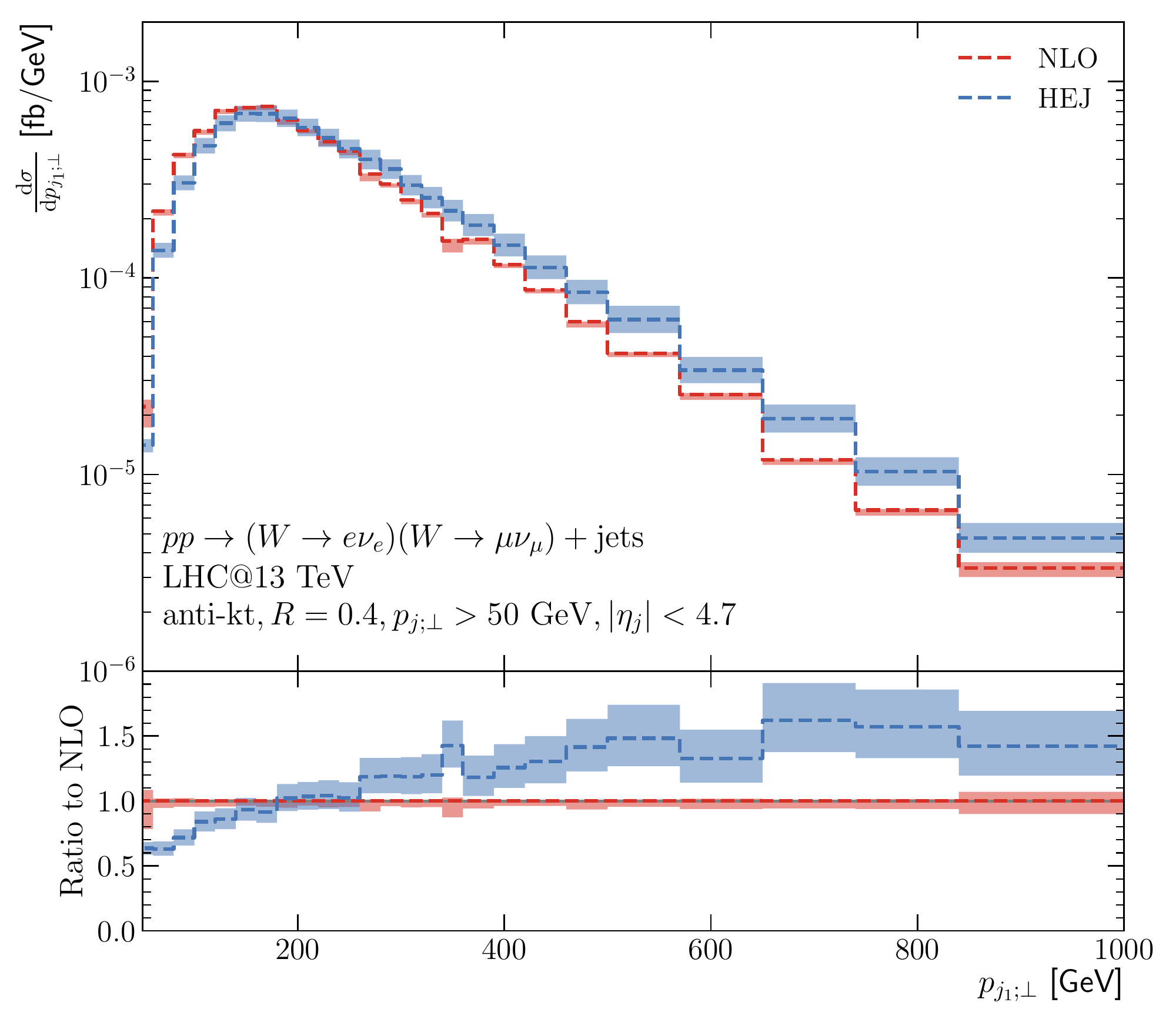}
		\caption{}
		\label{fig:m_j1j2_novbs}
	\end{subfigure}
	\begin{subfigure}{0.49\textwidth}
		\includegraphics[width=\textwidth]{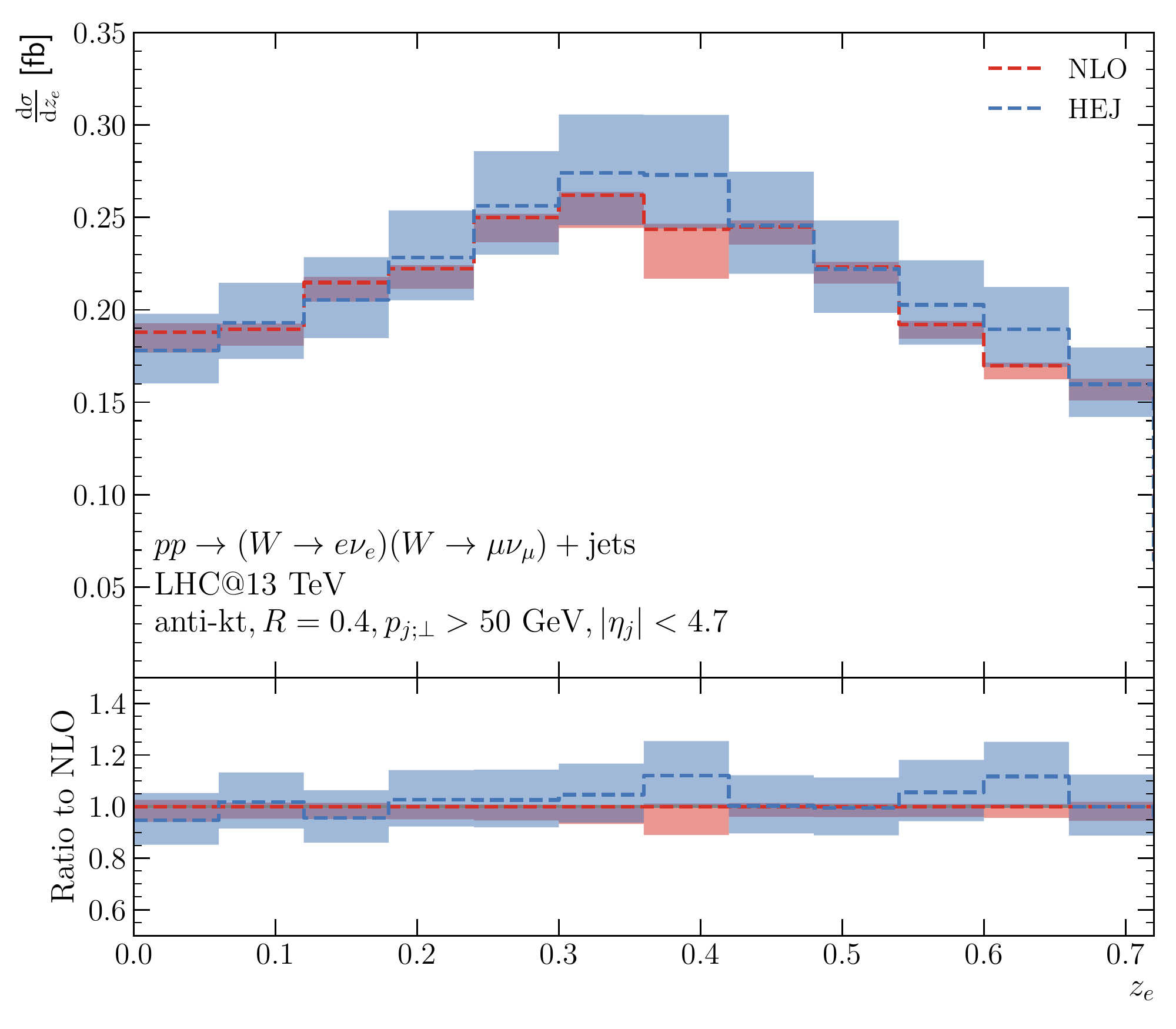}
		\caption{}
		\label{fig:m_j1j2_vbs}
	\end{subfigure}
	\caption{Differential distributions in the transverse momentum of the leading jet (a) and in the Zeppenfeld variable for the electron (b).}
	\label{fig:m_j1j2}
\end{figure}

From now on we focus on the results with the additional VBS cuts. In Figs.~\ref{fig:deltayjj} and~\ref{fig:m_j1j2} we show the comparison of the \HEJtwo and NLO calculations as a function of the pseudorapidity separation and invariant mass of the leading jets, the transverse momentum of the leading jet and the Zeppenfeld variable. One can see that the good agreement of the \HEJ and NLO cross sections does not apply to some distributions such as the transverse momentum of the leading jet and the invariant mass of the leading jets. In these cases the logarithmic corrections give predictions which are substantially larger than those predicted at NLO. On the other hand, the pseudorapidity separation of the leading jets and the Zeppenfeld variable show a close agreement between the two calculations, which indicates that these observables are less sensitive to the logarithmic corrections studied here.

\begin{table}
	\centering
	\begin{tabular}{|c|c|c|c|}
		\hline
		Cross Section (fb)& \emph{without} VBS cuts, $\sigma_{\rm incl}$ & \emph{with} VBS cuts, $\sigma_{\rm VBS} $ &
		$\sigma_{\rm VBS}$/$\sigma_{\rm incl}$   \\ \hline
		\HEJtwo $W^+W^+$ &$1.428 \pm 0.002 $&$0.1219 \pm 0.0004 $&$0.0854 \pm 0.0003 $\\
		NLO $W^+W^+$ &$1.41 \pm 0.05 $&$ 0.12 \pm 0.07 $&$0.08 \pm 0.02 $\\ \hline
		\HEJtwo $W^-W^-$ &$0.6586 \pm 0.0003 $&$0.0402\pm 0.0001 $&$0.0610 \pm 0.0002 $\\
		NLO $W^-W^-$ &$0.68 \pm 0.02 $&$ 0.04 \pm 0.01 $&$0.06 \pm 0.02 $\\ \hline
	\end{tabular}
	\caption{Cross section obtained at LO+LL accuracy using \HEJtwo compared with the result at NLO accuracy, both before and	after the additional VBS cuts given in the text.}
	\label{tab:crosssections}
\end{table}

\section{Conclusions}

We have presented a calculation of the process $pp\to e^\pm \nu_e \mu^\pm \nu_\mu+\ge2j$ at $\mathcal{O}(\alpha_W^4 \alpha_s^{k+2}\log^k(\hat s/p_t^2))$, focusing on typical cuts used in VBS analyses. These cuts select regions of the phase space where the high energy logarithms become important, and the \HEJtwo exclusive event generator provides a convenient framework to resum these contributions to all orders. While the inclusive cross sections obtained in this approach are very similar compared to those at NLO, significant differences can be observed in several distributions relevant for experimental studies, such as the exclusive jet rates, the invariant mass of the leading jets and the transverse momentum of the hardest jet. These results show that the logarithmic corrections proportional to $\alpha_s^k\log^k(\hat s/p_t^2)$ can be sizeable at the 13 TeV LHC and should be taken into account in order to accurately model the QCD contribution to vector boson scattering.

\section*{Acknowledgements}
We are grateful to our collaborators within \HEJ for useful discussions
throughout this project. The predictions presented in section \ref{sec:results} were produced using resources from PhenoGrid which is part of the GridPP Collaboration\cite{gridpp2006,gridpp2009}.

\paragraph{Funding information}
We are pleased to acknowledge funding from the UK
Science and Technology Facilities Council, the Royal Society, the ERC
Starting Grant 715049 ``QCDforfuture'' and the Marie
Sk{\l}odowska-Curie Innovative Training Network MCnetITN3 (grant
agreement no.~722104).

\bibliography{papers}

\nolinenumbers

\end{document}